\documentclass[twocolumn,showpacs,preprintnumbers,amsmath,amssymb]{revtex4}
\usepackage{mathrsfs}
\usepackage{amsfonts}
\usepackage{}

\usepackage{graphicx}
\usepackage{dcolumn}
\usepackage{bm}
\usepackage{amsmath,amsthm}

\begin{document}

\title{Determination of stabilizer states}

\author{Xia Wu$^{1,2}$}
\author{Ying-hui Yang$^{1}$}
\author{Yu-kun Wang$^{1}$}

\author{Qiao-yan Wen$^{1}$}
\author{Su-juan Qin$^{1}$}
\author{Fei Gao$^{1}$\footnote{Email: gaofei\_bupt@hotmail.com}}

\affiliation{
$^{1}$State Key Laboratory of Networking and Switching Technology, Beijing University of Posts and Telecommunications, Beijing, 100876, China\\
$^{2}$State Key Laboratory of Integrated Services Network, Xidian University,
Xi¡¯an 710071, China\\
}%

\date{\today}

\begin{abstract}
The determination of many special types of quantum states has been studied thoroughly, such as the generalized $|GHZ\rangle$ states, $|W\rangle$ states equivalent under stochastic local operations and classical communication and Dicke states. In this paper, we are going to study another special entanglement states which is  stabilizer states. The stabilizer states and their subset graph states   play an important role in quantum error correcting codes, multipartite purification and so on.
 We show that all $n$-qubit stabilizer states are uniquely determined (among arbitrary states, pure or mixed) by their reduced density matrices for systems which are the supports of $n$ independent generators of the corresponding stabilizer formalisms.

\end{abstract}

\maketitle


\section{\label{sec:level1}Introduction}

The recognition of quantum entanglement plays a key role in quantum information theory.   A quantum state is called entangled if
it is not classically correlated. The fact that local unitaries do not change entanglement is well known, that is two quantum states have the same entanglement if they can be get from each other by local unitary operators. This is different from local operations and classical communication, which may decrease entanglement. It is hard to quantify and describe the entanglement in the  multipartite quantum states.

 A basic question concerning quantum entanglement is that the whole information contained in a quantum state is dependent of its reduced states \cite{1,2,3,4}. That is whether a pure multi-qubit  quantum state is uniquely determined by its $k$-particle reduced density matrices is  concerned about, and at the same time  the smallest $k$  is desired. People call this  ``parts and whole''  problem. Much effort has been devoted to quantifying it so far.

 The  ``parts and whole'' problem was first proposed  in $2002$ by N.Linden \textsl{et al} \cite{1}, they proved that two-particle reduced density matrices are sufficient to uniquely determine almost every pure three-qubit state. They also shown that the set of just about half the parties contains all the information in almost all $n$-party pure states \cite{4}. In recent years much effort has been spent on studying quantum states with special forms and properties, such as stochastic local operations and classical communication (SLOCC) equivalent $|W_n\rangle$ states \cite{rana1,rana2,wu}, Dicke states \cite{dick}, stabilizer states \cite{zhou} and generalized $|GHZ_n\rangle$ states which are LU equivalent to some stabilizer states \cite{ghz1,ghz2}. In \cite{wu}, the present authors proved that SLOCC equivalent $|W_n\rangle$ states can be uniquely determined among arbitrary states by their $(n-1)$ bipartite reduced density matrices whose index set is correspond to a tree graph.  For $|GHZ_n\rangle$ states, it has been shown that there can be only $|GHZ_n\rangle$ states containing information at the $n$-particle level \cite{ghz1,ghz2}.

Stabilizer states play a central role in quantum correction and quantum computing \cite{stabilizer}, so the work we are doing in the determination of stabilizer states is dynamite. An $n$-qubit stabilizer state is the simultaneous eigenvector of an Abelian subgroup of the Pauli group which does not include $- \sigma _0^{ \otimes n}$. We call such an Abelian group as the stabilizer formalism of the stabilizer state.  For each stabilizer formalism, we can always find  $n$ independent commutative Pauli operators which generate this group, which are known as generators.
  As we know, the degrees of  irreducible $k$-particle correlation in $n$-qubit stabilizer states and generalized $|GHZ_n\rangle$  states have been obtained by Zhou in \cite{zhou}, and this reveals the distribution of multi-particle correlation of stabilizer states. But it is not given that by which reduced density matrices a stabilizer state is \emph{uniquely} determined. The main purpose of our paper is  to prove that $n$-qubit stabilizer states are uniquely determined (among arbitrary states, pure or mixed) by their reduced density matrices for systems which are the supports of $n$ independent generators of the corresponding stabilizer formalism. It is easy to show that reduced density matrices for systems which are ``smaller'' than the supports can not uniquely determine stabilizer states.

  The organization of this paper is as follows. In Sec. II we review the definition of stabilizer states and  introduce the binary generator matrix of stabilizer formalism which plays a crucial role in later proof. In Sec. III we first prove  the determination of stabilizer states among pure states which is much easier than treating with the mixed states. Then we prove our theorem holds among mixed states in Sec. IV. In Sec. V we summarize our results.

\section{\label{sec:level2}preliminaries: Stabilizer states and binary genenrator matrix}

Let $\mathcal{P}_{1}$ denotes the Pauli group on one qubit which is the multiplicative matrix group generated by the four Pauli matrices: $\sigma_{0},\sigma_{x},\sigma_{y}$  and $\sigma_{z}$. Pauli group on $n$ qubits, which is denoted by $\mathcal{P}_{n}$, is the $n$-fold tensor product of $\mathcal{P}_{1}$ with itself. Elements of $\mathcal{P}_{n}$ will be called Pauli operators. As we can see, an arbitrary Pauli operator can be  presented as the form
$\alpha {\sigma _{{\upsilon _1}}} \otimes {\sigma _{{\upsilon _2}}} \otimes  \cdots  \otimes {\sigma _{{\upsilon _n}}}$
,where $\alpha\in\{\pm1,\pm i\}$ is an overall phase factor and $\upsilon_{i}\in\{0,x,y,z\}$ for $i=1,\ldots,n$.
The Clifford group $\mathcal {C}_{1}$  on one qubit is the group of unitary operators which map the Pauli group $\mathcal{P}_{1}$ to itself under conjugation. Local Clifford group on $n$ qubits $\mathcal {C}_{n}$ consists of all $n$-fold tensor products of elements in  $\mathcal {C}_{1}$ \cite{p1c1}.

We denote $\mathcal {S}$ as the stabilizer on $n$ qubits. A stabilizer $\mathcal {S}$ is a subgroup of  $\mathcal{P}_{n}$ with all elements in which enjoy at least one common eigenvector with eigenvalue $1$. It has been proved that a subgroup $\mathcal {S}$ of $\mathcal{P}_{n}$ is a stabilizer on $n$ qubits if and only if $\mathcal {S}$ is an Abelian group which does not include $-\sigma_{0}^{\bigotimes n}$. The elements $M_{i}(i=1,\ldots,l)$ are called generators of  $\mathcal {S}$ if every element $M \in \mathcal {S}$ can be written as a product of $M_{i}$s, i.e., $M = M_1^{{x_1}}M_2^{{x_2}} \cdots M_l^{{x_l}}$ with $x_i\in \{0,1\}$. If $l=n$,  the elements in $\mathcal {S}$  will have only one common eigenvector with eigenvalue $1$. We call this unique common eigenvector as stabilizer state. In more details, a stabilizer state $\left| \psi  \right\rangle$ is the unique normalized state satisfying $M\left| \psi  \right\rangle=\left| \psi  \right\rangle$ for every $M \in \mathcal {S}$. It is easy to see that the density matrix of a stabilizer state $\left| \psi  \right\rangle$  is
\begin{equation}
\label{eq1}
\rho_{\mathcal {S}}=\left| \psi  \right\rangle \left\langle \psi  \right| = \frac{1}{{{2^n}}}\sum\limits_{M \in \mathcal {S}} M .
\end{equation}

Next, we will describe the stabilizer formalism in terms of the binary representation \cite{erjinzhi}. First, the connection between Pauli operators and binary algebra is obtained by employing the mapping Eq.(\ref{eq1})
\begin{equation}
\begin{array}{l}
 \alpha {\sigma _0} = \alpha {\sigma _{00}} \mapsto (0,0), \\
 \alpha {\sigma _x} = \alpha {\sigma _{01}} \mapsto (0,1), \\
 \alpha {\sigma _z} = \alpha {\sigma _{10}} \mapsto (1,0), \\
 \alpha {\sigma _y} = \alpha {\sigma _{11}} \mapsto (1,1). \\
 \end{array} \nonumber
\end{equation}
Therefore, we can introduce a mapping denoted by $\mathcal {B}$ which maps the elements of $\mathcal{P}_{n}$ to $2n$-dimensional binary vectors
 as follows:
\begin{equation}
\mathcal {B}:M = \alpha {\sigma _{{u_1}{v_1}}} \otimes  \cdots \otimes {\sigma _{{u_n}{v_n}}} = \alpha {\sigma _{(u,v)}} \mapsto  (u,v), \nonumber
\end{equation}
i.e., $\mathcal {B}(M)=(u,v)$, where $\alpha\in\{\pm1,\pm i\}$, $u=(u_1,\ldots,u_n), v=(v_1,\ldots,v_n) \in \mathbb{F}_2^{n}$. It has been shown that if the stabilizer formalism $\mathcal {S}$ of an $n$-qubit stabilizer state is generated by $n$ independent elements $M_1,\ldots,M_n$, then the set $\{\mathcal {B}(M_1),\ldots,\mathcal {B}(M_n)\}$ is a basis of $\mathcal {B}(\mathcal {S})$. For  stabilizer formalism  with $n$ independent generators, we present binary space $\mathcal {B}(\mathcal {S})$ in terms of a generator matrix $S$ which is an $2n\times n$ matrix over $\mathbb{F}_{2}$ and the columns form a basis of $\mathcal {B}(\mathcal {S})$:
\begin{equation}
S=[\mathcal {B}(M_{1})|\ldots|\mathcal {B}(M_{n})], \nonumber
\end{equation}
where $M_1,\ldots,M_n$ are generators of $\mathcal {S}$.

As an important subclass of stabilizer states, graph states have got a lot of attentions \cite{graph state}.
Formally, a graph is a pair $G=(V,E)$ of sets satisfying $E\subseteq [V]^{2}$, where $V=\{1,2,\cdots,N\}$ and the elements of $E$ are $2$-element subsets of $V$. We call the elements in $V$ as the vertices  and the elements in $E$ as  edges. In our paper, we will only consider undirected simple graphs, which have no multiple edges and no loops (a loop is an edge of the form $(i,i)$, where $i\in V$). Two vertices $i,j\in V$ are called adjacent vertices if $(i,j)\in E$.  The adjacency relations between the $n$ vertices of a simple graph $G$ are described by an adjacency matrix $\theta(G)=\theta$ which is the symmetric binary $n\times n$ matrix defined by: $\theta_{ij}=1$ if $(i,j)\in E$,  $\theta_{ij}=0$ otherwise.
 For a given  simple graph on $n$ vertices, the adjacency matrix of which is denoted by $\theta$, one can define $n$ commutative Pauli operators with special forms as follows:
\begin{equation}
{K_s} = \sigma _x^s\prod\limits_{t = 1}^n {{{(\sigma _z^t)}^{{\theta _{st}}}}}, \nonumber
\end{equation}
where the index $s$  of Pauli operators means: in the tensor product, $\sigma_{x}$ is on the $s$-th position.
As we can see, the set which is generated by $K_1,\ldots,K_n$ is a stabilizer, and we call the corresponding stabilizer state as the  graph state on $n$ vertices which is denoted as  $|G_{n}\rangle$. Note that the generator matrix of $|G_{n}\rangle$ is $S = {\left[ {\begin{array}{*{20}{c}}
   \theta  & I  \\\end{array}} \right]^T}$.

It has been proved that there is a common way to describe graph states in terms of quadratic forms \cite{graph state2}. For a given  simple graph with the adjacency matrix $\theta$, it can be associated with the following quadratic form:
\begin{equation*}
{f_\theta }(x) = \sum\limits_{\scriptstyle 1 \le s,t \le n \atop
  \scriptstyle s < t} {{\theta _{st}}{x_s}{x_t}}  \in {\mathbb{F}_2}.
\end{equation*}
By using the function $f_{\theta}$, we can expansion the $n$-qubit graph state in the computational basis as follows:
\begin{equation}
\left| {{G_n}} \right\rangle  = \frac{1}{{\sqrt {{2^n}} }}\sum\limits_{x \in \mathbb{F}_2^n} {{{( - 1)}^{{f_\theta }(x)}}\left| x \right\rangle }.
\end{equation}

Next, let us recall the definition of \textsl{support} of a Pauli operator. Let an arbitrary Pauli operator $M = \alpha {\sigma _{{u_1}{v_1}}} \otimes  \cdots \otimes {\sigma _{{u_n}{v_n}}}$, the \textsl{support} of $M$ is the set
\begin{equation}
\textrm{supp}(M)=\{j\in\{1,\ldots,n\}|(u_{j},v_j)\neq(0,0)\}. \nonumber
\end{equation}
Note that $\textrm{supp}(M)$ contains those $j$ satisfying the $j$th tensor factor of $M$ differs from  $\sigma_{0}$.
Secondly, for any $\omega\subseteq \{1,\ldots,n\}$ the Pauli operator $M_{\omega} \in \mathcal{P}_{|\omega|}$ of $M = \alpha {\sigma _{{u_1}{v_1}}} \otimes  \cdots \otimes {\sigma _{{u_n}{v_n}}}$ is defined by
\begin{equation}
{M_\omega } = \alpha { \otimes _{j \in \omega }}{\sigma _{{u_j}{v_j}}}. \nonumber
\end{equation}
Thus $M_\omega$ is the tensor product of the $j$-th (all $j\in \omega$) tensor factor of $M$.

It can be  easily obtained by Eq.(\ref{eq1}) that, for an arbitrary stabilizer state $\rho_{\mathcal {S}}$, tracing out all qubits of $\rho_{\mathcal {S}}$ outside a set $\omega$ yields a state $\rho_{\mathcal {S}}^{\omega}$, which is equal to
\begin{equation}
\rho _\mathcal {S}^\omega  = \frac{1}{{{2^{|\omega |}}}}\sum\limits_{M \in S,\textrm{supp}(M) \subseteq \omega } {{M_\omega }}.
\end{equation}
$\rho _\mathcal {S}^\omega$ is exactly the reduced density matrix for system $\omega$ of the stabilizer state.

\section{\label{sec:level3}AMONG PURE STATES}

 M. Van den Nest \textsl{et al}. \cite{p1c1} have proved that every stabilizer state is local Clifford (LC) equivalent to some graph state,  where local Clifford group is the subgroup of local unitary group. It's obvious that  the determination of  quantum states is invariant under local unitary actions, that is if an $n$-qubit pure state is uniquely determined by its $k$ reduced density matrices, then its LU equivalent states are also uniquely determined by their $k$ reduced density matrices. Therefore we only need to deal with the case of graph states, i.e., the determination of graph states. Such a discussion is sufficient and complete.
\footnotemark\footnotetext{Here, we use that  if an $n$-qubit pure state is uniquely determined by its $k$ reduced density matrices and a subset of the reduced density matrices set is enough, then its LU equivalent states are also uniquely determined by the same subset of the reduced density matrices. We conjecture this proposition is correct, but we have not proven it and cannot find an  counterexample or  if it has been proved in some works. If someone get anything useful, you are welcome to contact us.}

We have presented the reduced density matrix of a stabilizer state, how the reduced density matrix of an arbitrary pure state can be expressed?  We denote a general $n$-qubit pure state as
\begin{equation}
\left| {{\varphi _n}} \right\rangle  = \sum\limits_{{i_1}, \ldots ,{i_n} = 0}^1 {{a_{{i_1} \ldots {i_n}}}\left| {{i_1} \ldots {i_n}} \right\rangle }  = \sum\limits_{i \in \mathbb{F}_2^n} {{a_i}\left| i \right\rangle }. \nonumber
\end{equation}
The reduced density matrix for system $\omega\subseteq \{1,\ldots,n\}$ is noted as $\rho _{{\varphi _n}}^\omega $. It is known to us that the reduced density matrix for system $\omega$ is matrix of $2^{|\omega|}\times2^{|\omega|}$, so we are going to calculate the coefficient of $\left| {{i_\omega }} \right\rangle \left\langle {{j_\omega }} \right|$ which is correspond to the element at the ($D(i_{\omega})+1$)-th row and ($D(j_{\omega})+1$)-th column of $\rho _{{\varphi _n}}^\omega $, where if $\omega=\{\omega_{1},\ldots,\omega_{h}\}$, we denote ${i_\omega } = ({i_{{\omega _1}}} \cdots {i_{{\omega _h}}}) $, ${j_\omega } = ({j_{{\omega _1}}} \cdots {j_{{\omega _h}}}) $ which are all binary vectors in $ \mathbb{F}_2^{\left| \omega  \right|}$; $D(i_{\omega})$, $D(j_{\omega})$ are respectively the decimal numbers of the binary vectors $i_\omega$ and $j_\omega$. Then according to the analysis in \cite{wu} we have the coefficient of $\left| {{i_\omega }} \right\rangle \left\langle {{j_\omega }} \right|$ is
\begin{equation}
\label{yuehuazhen}
\sum\limits_{{i_{\bar \omega }} \in \mathbb{F}_2^{|\bar \omega |}} {{a_{{i_\omega }{i_{\bar \omega }}}} \cdot {{ a^{*}}_{{j_\omega }{i_{\bar \omega }}}}},
\end{equation}
where $\bar \omega$ is the complement set of $\omega$ in $\{1,\ldots,n\}$.

Before we deal with the determination of graph states, we first state  a lemma about the property of Pauli operators which can be easily obtained from the properties of the $4$ Pauli matrices.

\textit{Lemma.}
Let an arbitrary Pauli operator is $M =(a_{i,j})_{2^n\times2^n}= \alpha {\sigma _{{u_1}{v_1}}} \otimes  \cdots \otimes {\sigma _{{u_n}{v_n}}}=\alpha\sigma _{(u,v)}$, and the support of $M$ is $\textrm{supp}(M) = \omega   \subseteq \{ 1, \ldots ,n\} $. ${M_\omega }$ is  denoted as  ${M_\omega } = {({b_{i,j}})_{{2^{|\omega |}} \times {2^{|\omega |}}}}$. Then we have

 (i) For any $({i_1} \cdots {i_n}) \in \mathbb{F}_2^{n}$, ${a_{{i_1} \cdots {i_n},({i_1} \cdots {i_n} + v)}}$is the only element in $M$ which is not equal to zero in the  row and column it presents;

(ii) For any ${i_\omega } \in \mathbb{F}_2^{|\omega |}$,we have
\begin{equation}
{b_{{i_\omega },({i_\omega } + {v_\omega })}} = {a_{{i_\omega }{i_{\bar \omega }},({i_\omega }{i_{\bar \omega }} + v)}},
 \end{equation}
 where ${i_{\bar \omega }}$ is an arbitrary vector in space  $ \mathbb{F}_2^{|\bar \omega |}$ and the "+" is performed in the finite field $\mathbb{F}_{2}$.

It has been noted that the $n$-qubit $|GHZ_n\rangle$ state is also a stabilizer state, which has a stabilizer formalism $\mathcal {S}$ generated by the $n$ elements
\begin{equation}
\begin{array}{l}
 {\sigma _x} \otimes {\sigma _x} \otimes {\sigma _x} \otimes {\sigma _x} \otimes  \cdots  \otimes {\sigma _x}, \\
 {\sigma _z} \otimes {\sigma _z} \otimes {\sigma _0} \otimes {\sigma _0} \otimes  \cdots  \otimes {\sigma _0}, \\
 {\sigma _0} \otimes {\sigma _z} \otimes {\sigma _z} \otimes {\sigma _0} \otimes  \cdots  \otimes {\sigma _0}, \\
 {\sigma _0} \otimes {\sigma _0} \otimes {\sigma _z} \otimes {\sigma _z} \otimes  \cdots  \otimes {\sigma _0}, \\
  \cdots  \\
 \end{array} \nonumber
\end{equation}
It is LC equivalent to the graph state with star graph. As we can see, the $|GHZ_n\rangle$ state is the one containing information at the $n$-party level, and at the same time, the maximum support of its generators  is always $\{1,\ldots,n\}$. Therefore, we can conjecture that there is some relationship between the determination of  stabilizer states and their generators of stabilizer formalism.

First of all, we are going to deal with the determination of graph states among pure states for easier understanding.

 \textit{Theorem 1.}
 Let $|G_n\rangle$  be a  graph state and   an arbitrary generating set of its stabilizer formalism is denoted as $\{M_1,\ldots,M_n\}$.
Then among pure states, the graph state is uniquely determined by its reduced density matrices set of
\begin{equation}
R = \left\{ {{\rho ^{{\rm{supp}}({M_s})}}|s = 1, \ldots ,n} \right\} \nonumber
\end{equation}

\begin{proof}
Let $S$ be the generator matrix of  $|G_n\rangle$, i.e., $S=[\mathcal {B}(M_{1})|\ldots|\mathcal {B}(M_{n})]\in \mathbb{F}_{2}^{2n\times n}$, and denote $S = \left( {\begin{array}{*{20}{c}}
   {{S_z}}  \\
   {{S_x}}  \\
\end{array}} \right)$ , where ${S_z},{S_x}$ are all $n\times n$ blocks. Next, we express ${S_x}$ with column vectors, that is let ${S_x} = \left( {\begin{array}{*{20}{c}}
   {r_1^T} &  \cdots  & {r_n^T}  \\
\end{array}} \right)
$, where $r_s^T$ is the $s$-th column of ${S_x}$, and as we can see ${r_s} \in \mathbb{F}_2^n$ is the last $n$ elements of $\mathcal {B}({M_s})$, $s=1,\ldots,n$. From the property of graph states, we can get matrix ${S_x}$ has full rank, then ${r_1}, \ldots ,{r_n}$ are a tuple of independent vectors, they are one of the linear independent basis of space $\mathbb{F}_2^n$.

Let the support of generator ${M_s}$ is $\textrm{supp}(M)=\omega_s$, then we can get ${M_{{\omega _s}}}$ from ${M_s}$ easily. Furthermore, binary representation $\mathcal {B}({M_{{\omega _s}}})$ is the vector belongs to space $F_2^{2|{\omega _s}|}$, and we denote a new vector as ${r_{{\omega _s}}}$ which is constituted by the  last $\left| {{\omega _s}} \right|$ elements. It is easy to see that ${r_{{\omega _s}}}$ is also composed of the bits in the position ${\omega _s}$ of ${r_s}$.

 To prove the theorem, we will show that for an $n$-qubit pure state $\left| {{\varphi _n}} \right\rangle  =  \sum\limits_{i \in \mathbb{F}_2^n} {{a_i}\left| i \right\rangle }$, if the reduced density matrices satisfying $\rho _\varphi ^{\omega_s} = \rho _G^{\omega_s}$, $s = 1, \ldots ,n$, then $\left| {{\varphi _n}} \right\rangle  = \left| {{G_n}} \right\rangle$.

According to the preliminaries we can get, for an arbitrary graph state,the reduced density matrix for system ${\omega _s}$, which is the support of the generator $M_s$, enjoys very simple form:
\begin{equation}
\rho _G^{{\omega _s}} = \frac{1}{{{2^{\left| {{\omega _s}} \right|}}}}\sum\limits_{M \in \mathcal {S},\textrm{supp}(M) \subseteq {\omega _s}} M  = \frac{1}{{{2^{\left| {{\omega _s}} \right|}}}}(I + {M_{{\omega _s}}} +  \cdots )\nonumber
\end{equation}
Then we have the coefficient of $\left| {0 \cdots 0} \right\rangle \left\langle {{r_{{\omega _s}}}} \right|$ with  $\rho _G^{{\omega _s}}$ is equal to the coefficient of $\left| {0 \cdots 0} \right\rangle \left\langle {{r_{{\omega _s}}}} \right|$ with $\frac{1}{{{2^{\left| {{\omega _s}} \right|}}}}{M_{{\omega _s}}}$ (This because $\{ M|\textrm{supp}(M) \subseteq {\omega _s}\}$ is also a stabilizer set without ${\sigma _z} \otimes \cdots \otimes {\sigma _z}$). However, in the light of Lemma (ii), we have the coefficient of $\left| {0 \cdots 0{i_{{{\bar \omega }_s}}}} \right\rangle \left\langle {{r_{{\omega _s}}}{i_{{{\bar \omega }_s}}}} \right|$ with ${M_s}$ is also equal to the coefficient of $\left| {0 \cdots 0} \right\rangle \left\langle {{r_{{\omega _s}}}} \right|$ with ${M_{{\omega _s}}}$, where ${i_{{{\bar \omega }_s}}}$ is an arbitrary vector belongs to the space $\mathbb{F}_2^{|{{\bar \omega }_s}|}$. Furthermore, for any binary vector ${{i_{{\omega _s}}}}$, we have the coefficient of $\left| {{i_{{\omega _s}}}} \right\rangle \left\langle {{i_{{\omega _s}}} + {r_{{\omega _s}}}} \right|$ with $\rho _G^{{\omega _s}}$ is equal to the coefficient of
\begin{equation}
\left| {{i_{{\omega _s}}}{i_{{{\bar \omega }_s}}}} \right\rangle \left\langle {({i_{{\omega _s}}} + {r_{{\omega _s}}}){i_{{{\bar \omega }_s}}}} \right| = \left| {{i_{{\omega _s}}}{i_{{{\bar \omega }_s}}}} \right\rangle \left\langle {{i_{{\omega _s}}}{i_{{{\bar \omega }_s}}} + {r_s}} \right| \nonumber
\end{equation}
with $\frac{1}{{{2^{\left| {{\omega _s}} \right|}}}}{M_s}$, where any ${i_{{{\bar \omega }_s}}} \in \mathbb{F}_2^{\left| {{{\bar \omega }_s}} \right|}$. From Eq.(1) and the computational basis representation of graph states in Eq.(6), we can get the coefficient  of arbitrary $|i\rangle\langle j|$ with ${\rho _G}$ is
\begin{equation}
{( - 1)^{f(i) + f(j)}} \cdot \frac{1}{{{2^n}}}, \nonumber
\end{equation}
where $i,j\in \mathbb{F}_{2}^n$.

Next we only need to focus on the coefficient of $\left| {{i_{{\omega _s}}}} \right\rangle \left\langle {{i_{{\omega _s}}} + {r_{{\omega _s}}}} \right|$ with $\rho _\varphi ^{{\omega _s}}$ . As $\rho _\varphi ^{{\omega _s}} = \rho _G^{{\omega _s}}$ and according to Eq.(\ref{yuehuazhen}), we have
\begin{equation}
\sum\limits_{{i_{{{\bar \omega }_s}}} \in \mathbb{F}_2^{|{{\bar \omega }_s}|}} {{a_{{i_{{\omega _s}}}{i_{{{\bar \omega }_s}}}}} \cdot {{ a^*}_{{i_{{\omega _s}}}{i_{{{\bar \omega }_s}}} + {r_s}}}}  = \frac{1}{{{2^{|{\omega _s}|}}}}{( - 1)^{f({i_{{\omega _s}}}{i_{{{\bar \omega }_s}}}) + f({i_{{\omega _s}}}{i_{{{\bar \omega }_s}}} + {r_s})}} \nonumber
\end{equation}

Since the diagonal elements of $\rho _G^{{\omega _s}}$ are all equal to $\frac{1}{{{2^{\left| {{\omega _s}} \right|}}}}$, the elements in the   diagonal position $\left| {{i_{{\omega _s}}}} \right\rangle \left\langle {{i_{{\omega _s}}}} \right|$ with $\rho _\varphi ^{{\omega _s}}$ satisfy:
\begin{equation}
\sum\limits_{{i_{{{\bar \omega }_s}}} \in F_2^{|{\omega _s}|}} {{a_{{i_{{\omega _s}}}{i_{{{\bar \omega }_s}}}}} \cdot {{a^*}_{{i_{{\omega _s}}}{i_{{{\bar \omega }_s}}}}}}  = \frac{1}{{{2^{\left| {{\omega _s}} \right|}}}} \nonumber
\end{equation}
According to the Schwarz inequality, we have
\begin{equation}
\label{xuyaojianhua}
{a_{{i_{{\omega _s}}}{i_{{{\bar \omega }_s}}}}} = {( - 1)^{f({i_{{\omega _s}}}{i_{{{\bar \omega }_s}}}) + f({i_{{\omega _s}}}{i_{{{\bar \omega }_s}}} + {r_s})}} \cdot {a_{{i_{{\omega _s}}}{i_{{{\bar \omega }_s}}} + {r_s}}},
\end{equation}
where ${i_{{\omega _s}}}, {i_{{{\bar \omega }_s}}}$ are two arbitrary vectors belong to space $\mathbb{F}_2^{|{\omega _s}|}, \mathbb{F}_2^{|{{\bar \omega }_s}|}$ respectively.
Thus we can simplify the above Eq.(\ref{xuyaojianhua}) to the form as follows:
\begin{equation}
\label{eq22}
{( - 1)^{f(r)}} \cdot {a_r} = {( - 1)^{f(r + {r_s})}} \cdot {a_{r + {r_s}}},
\end{equation}
where any $r\in \mathbb{F}_{2}^n$, and $s = 1, \ldots ,n$.

As ${r_1}, \ldots ,{r_n}$ is a tuple of basis with space $\mathbb{F}_2^n$, for any vector $r \in \mathbb{F}_2^n$, we have $r = {x_1}{r_1} +  \cdots  + {x_n}{r_n}$, where $x_i \in \{0,1\}$. Furthermore, we can get from Eq.(\ref{eq22}) that

\begin{equation}
\begin{split}
\label{eq23}
 &{( - 1)^{f(r)}} \cdot {a_r} \\
& = {( - 1)^{f(r + {x_n}{r_n})}} \cdot {a_{r + {x_n}{r_n}}}  \\
  &= {( - 1)^{f({x_1}{r_1} +  \cdots  + {x_{n - 1}}{r_{n - 1}})}} \cdot {a_{{x_1}{r_1} +  \cdots  + {x_{n - 1}}{r_{n - 1}}}} \\
  &= \cdots \\
  &= {( - 1)^{f({x_1}{r_1})}} \cdot {a_{{x_1}{r_1}}} \\
  &= {( - 1)^{f({x_1}{r_1} + {x_1}{r_1})}} \cdot {a_{{x_1}{r_1} + {x_1}{r_1}}} \\
  &= {( - 1)^{f(0)}} \cdot {a_0} = {a_0} \\
 \end{split} 
\end{equation}
Then for any vector $r\in \mathbb{F}_{2}^n$, we have
\begin{equation}
{a_r} = {( - 1)^{f(r)}} \cdot {a_0}. \nonumber
\end{equation}

Therefore, we can get
\begin{eqnarray}
\begin{split}
 \left| {{\varphi _n}} \right\rangle  & = \sum\limits_{r \in \mathbb{F}_2^n} {{a_r}\left| r \right\rangle }  = \sum\limits_{r \in \mathbb{F}_2^n} {{{( - 1)}^{f(r)}}{a_0}\left| r \right\rangle } \\
& = {a_0}\sqrt {{2^n}} \left| {{G_n}} \right\rangle = \left| {{G_n}} \right\rangle . \\
 \end{split} \nonumber
\end{eqnarray}

End the proof.

\end{proof}

From the proof above, it can be seen that the binary presentation of Pauli operators playes a crucial role. We do not need to examine all of the nonzero elements in reduced density matrices $\rho _G^{{\omega _s}}$, since only  the nonzero elements in ${M_{{\omega _s}}}$ are enough .

\section{\label{sec:level4}AMONG arbitrary STATES}

In this section we show that the determination of stabilizer states among arbitrary states is the same as the determination among pure states. It is known to us that, the proof of determination among arbitrary states is much more difficult than the proof among pure states, the reason is that when we  deal with arbitrary states, ${2^{n - 1}}({2^n} + 1)$ variables should be determined, which is largely more than $n$ variables when we deal with pure states. Fortunately, there are many good properties with density matrix which will be used in our proof, such as the trace of a density matrix is $1$ and all principle minors of a density matrix are non-negative.

 \textit{Theorem 2.}
 Let $|G_n\rangle$  be an $n$-qubit graph state and   an arbitrary generating set of its stabilizer formalism is denoted as $\{M_1,\ldots,M_n\}$.
Then among arbitrary states, the graph state is determined by its reduced density matrices set of
\begin{equation*}
R = \left\{ {{\rho ^{{\rm{supp}}({M_s})}}|s = 1, \ldots ,n} \right\}
\end{equation*}

\begin{proof}

We can write a general (possibly mixed) $n$-qubit density matrix in standard computational basis as
\begin{equation*}
{\rho _A} = \sum\limits_{i,j \in \mathbb{F}_2^n} {{b_{ij}}\left| i \right\rangle \left\langle j \right|}
\end{equation*}
What we are going to prove is that if $\rho _{\rm{A}}^{{\textrm{supp}}({M_s})} = \rho _G^{{{\textrm{supp}}}({M_s})}$ for $s=1,\ldots,n$, then ${\rho _A} = {\rho _G}$.

From the analysis in Theorem $1$, for any ${i_{{\omega _s}}} \in \mathbb{F}_2^{|{\omega _s}|}$, and the coefficient of $\left| {{i_{{\omega _s}}}} \right\rangle \left\langle {{r_{{\omega _s}}} + {i_{{\omega _s}}}} \right|$, we have
\begin{equation}
\label{eq28}
\sum\limits_{{i_{{{\bar \omega }_s}}} \in \mathbb{F}_2^{|{{\bar \omega }_s}|}} {{b_{({i_{{\omega _s}}}{i_{{{\bar \omega }_s}}})({i_{{\omega _s}}}{i_{{{\bar \omega }_s}}} + {r_s})}}}  = \frac{1}{{{2^{|{\omega _s}|}}}}{( - 1)^{f({i_{{\omega _s}}}{i_{{{\bar \omega }_s}}}) + f({i_{{\omega _s}}}{i_{{{\bar \omega }_s}}} + {r_s})}}
\end{equation}
For the coefficient of $\left| {{i_{{\omega _s}}}} \right\rangle \left\langle {{i_{{\omega _s}}}} \right|$, we have
\begin{equation}
\label{eq29}
\sum\limits_{{i_{{{\bar \omega }_s}}} \in \mathbb{F}_2^{|{{\bar \omega }_s}|}} {{b_{({i_{{\omega _s}}}{i_{{{\bar \omega }_s}}})({i_{{\omega _s}}}{i_{{{\bar \omega }_s}}})}}}  = \frac{1}{{{2^{|{\omega _s}|}}}}.
\end{equation}
Then for  $\forall  r \in \mathbb{F}_2^n$ we can get the following conclusions which is proved
in Appendix A:
\begin{equation}
\label{eq27}
\left| {{b_{(r)(r + {r_s})}}} \right| = \sqrt {{b_{(r)(r)}}{b_{(r + {r_s})(r + {r_s})}}},
\end{equation}
and
\begin{equation}
{b_{(r)(r)}} = {b_{(r + {r_s})(r + {r_s})}}. \nonumber
\end{equation}
As $r = {x_1}{r_1} +  \cdots  + {x_n}{r_n}$,  $x_i \in \{0,1\}$, following the similar analysis with Eq.(\ref{eq23}), we can get ${b_{(r)(r)}}={b_{(0)(0)}}$.
The trace with the density matrix is $1$ implies $2^n\cdot{b_{(0)(0)}} = 1$,
thus  we have ${b_{(r)(r)}} $ is equal to $\frac{1}{{{2^n}}}$ and $\left| {{b_{(r)(r + {r_s})}}} \right|$ is also equal to $\frac{1}{{{2^n}}}$ according to Eq.(\ref{eq27}).
Thus from Eq.(\ref{eq28}), for any $ {i_{{\omega _s}}} \in \mathbb{F}_2^{\left| {{\omega _s}} \right|}$ and $ {i_{{{\bar \omega }_s}}} \in \mathbb{F}_2^{\left| {{{\bar \omega }_s}} \right|}$,it follows
\begin{equation}
{b_{({i_{{\omega _s}}}{i_{{{\bar \omega }_s}}})({i_{{\omega _s}}}{i_{{{\bar \omega }_s}}} + {r_s})}} = {( - 1)^{f({i_{{\omega _s}}}{i_{{{\bar \omega }_s}}}) + f({i_{{\omega _s}}}{i_{{{\bar \omega }_s}}} + {r_s})}} \cdot \frac{1}{{{2^n}}},\nonumber
\end{equation}
that is for any $ r \in \mathbb{F}_2^n$, we can get
\begin{equation}
\label{eq34}
{b_{(r)(r + {r_s})}} = {( - 1)^{f(r) + f(r + {r_s})}} \cdot \frac{1}{{{2^n}}}
\end{equation}

 Collecting the results above, we should know that we have got:  the diagonal elements with $\rho _A$ are all $\frac{1}{{{2^n}}}$ and some of the non-diagonal elements which are shown in Eq.(\ref{eq34}). Next, the only remaining task is to  prove
 \begin{equation}
 {b_{(i)(j)}} = {( - 1)^{f(i) + f(j)}} \cdot \frac{1}{{{2^n}}},\nonumber
 \end{equation}
 for $\forall i,j \in \mathbb{F}_2^n$. Before we examine the value of ${b_{(i)(j)}}$,  we first prove that for  any $ j \in \mathbb{F}_2^n$, it follows ${b_{(0)(j)}} = {( - 1)^{ f(j)}} \cdot \frac{1}{{{2^n}}}$.

Since $\left\{ {{r_1}, \ldots ,{r_n}} \right\}$ is a basis of space $\mathbb{F}_2^n$, any binary vector $ j \in \mathbb{F}_2^n$ can be denoted as $j = {x_1}{r_1} +  \cdots  + {x_n}{r_n}$, where $x_i \in \{0,1\}$. We may assume that ${x_{{k_1}}} = {x_{{k_2}}} =  \cdots  = {x_{{k_m}}} = 1$ ; $x_k=0$ elsewhere, that is we assume $j = {r_{{k_1}}} +  \ldots  + {r_{{k_m}}}$.  Owing to the principle minors of a density matrix are non-negative, let us consider the following principle minor consisting of the rows and columns $\left( {0 \ldots 0} \right),r_{{k_1}}, r_{{k_1}}+r_{{k_2}}$:
\begin{widetext}
\[\frac{1}{{{2^{3n}}}}\left| {\begin{array}{*{20}{c}}
  \vspace{0.20cm} 1 & {{{( - 1)}^{f({r_{{k_1}}})}}} & {{2^n}\cdot{b_{(0)({r_{{k_1}}} + {r_{{k_2}}})}}}  \\\vspace{0.20cm}
 \vspace{0.20cm}  {{{( - 1)}^{f({r_{{k_1}}})}}} & 1 & {{{( - 1)}^{f({r_{{k_1}}}) + f({r_{{k_1}}} + {r_{{k_2}}})}}}  \\ \vspace{0.20cm}
 \vspace{0.20cm}  {{2^n}\cdot{{\bar b}_{(0)({r_{{k_1}}} + {r_{{k_2}}})}}} & {{{( - 1)}^{f({r_{{k_1}}}) + f({r_{{k_1}}} + {r_{{k_2}}})}}} & 1  \\
\end{array}} \right|\]
\end{widetext}
The value of this determinant is $- \frac{1}{{{2^n}}}{\left| {{b_{\left( 0 \right)\left( {{r_{{k_1}}} + {r_{{k_1}}}} \right)}} - {{( - 1)}^{f({r_{{k_1}}} + {r_{{k_1}}})}} \cdot \frac{1}{{{2^n}}}} \right|^2}$.
Since this should be non-negative, it follows
\begin{equation}
 {{b_{\left( 0 \right)\left( {{r_{{k_1}}} + {r_{{k_1}}}} \right)}} = {{( - 1)}^{f({r_{{k_1}}} + {r_{{k_1}}})}} \cdot \frac{1}{{{2^n}}}} \nonumber
\end{equation}

Next, let us consider the  principle minor consisting of the rows and columns $\left( {0 \ldots 0} \right),{r_{{k_1}}} + {r_{{k_2}}},{r_{{k_1}}} + {r_{{k_2}}} + {r_{{k_3}}}$, we can easily get
\begin{equation}
{b_{\left( 0 \right)\left( {{r_{{k_1}}} + {r_{{k_2}}} + {r_{{k_3}}}} \right)}} = {\left( { - 1} \right)^{ f({r_{{k_1}}} + {r_{{k_2}}} + {r_{{k_3}}})}} \cdot \frac{1}{{{2^n}}} \nonumber
\end{equation}
\[\vdots\]
Therefore,  for any $j \in \mathbb{F}_2^n$, we can get
\begin{equation}
{b_{(0)(j)}} = {( - 1)^{ f(j)}} \cdot \frac{1}{{{2^n}}}. \nonumber
\end{equation}

Finally, to complete the proof we only need to examine the value of ${b_{(i)(j)}}$, where  $i\neq 0$, $j\neq 0$. Let us consider the principle minor consisting of the rows and columns $\left( {0 \ldots 0} \right),i,j$,
and we get the  value of the determinant is $- \frac{1}{{{2^n}}}{\left| {{b_{(i)(j)}} - {{( - 1)}^{f(i) + f(j)}} \cdot \frac{1}{{{2^n}}}} \right|^2}$.
Since this should be non-negative, it follows
\begin{equation}
{b_{(i)(j)}} = {( - 1)^{f(i) + f(j)}} \cdot \frac{1}{{{2^n}}}. \nonumber
\end{equation}
This completes the proof.
\end{proof}

 As we can see in our proof, we do not need to care about which is bigger in these two numbers $D(k)$ and $D(k+r_s)$, since the determinant is not changed by removing of just one row and one column.
As a result of the above theorem, one can quickly get the following corallary.

 \textit{Corollary.}
Let $|G_n\rangle$  be an $n$-qubit graph state and   an arbitrary generating set of its stabilizer formalism is denoted as $\{M_1,\ldots,M_n\}$. One can always find out the set $V' \subseteq \{1,\ldots,n\}$ that the element $s$ belongs to $V'$ if $\textrm{supp}(M_s)\nsubseteq \textrm{supp}(M_t)$ for $\forall t \in V'\setminus s$.
Then among arbitrary states, the graph state is determined by its reduced density matrices set of
\begin{equation}
R = \{\rho^{\textrm{supp}(M_s)}|s \in V' \} \nonumber
\end{equation}

As an example, for the $4-$qubit graph state $|G_4\rangle$ which is generated by the $4$ elements:
\begin{equation}
\label{eq40}
\begin{array}{l}
 {\sigma _x} \otimes {\sigma _z} \otimes {\sigma _0} \otimes {\sigma _0}, \\
 {\sigma _z} \otimes {\sigma _x} \otimes {\sigma _z} \otimes {\sigma _0}, \\
 {\sigma _0} \otimes {\sigma _z} \otimes {\sigma _x} \otimes {\sigma _z}, \\
 {\sigma _0} \otimes {\sigma _0} \otimes {\sigma _z} \otimes {\sigma _x}. \\
 \end{array}
\end{equation}
Then $|G_4\rangle$ is uniquely determined among arbitrary states by its reduced density matrices $\{{\rho ^{123}},{\rho ^{234}}\}$, $\{{\rho ^{123}},{\rho ^{134}}\}$, $\{{\rho ^{124}},{\rho ^{234}}\}$ and $\{{\rho ^{124}},{\rho ^{134}}\}$.
\section{\label{sec:level5}CONCLUSION}
The study of the determination of stabilizer states is, undoubtedly, important and even necessary to ``parts and whole'' problem. Although the distribution of multiparty correlations with $n$-qubit stabilizer states has been revealed, which reduced density matrices can uniquely determine the stabilizer states have not been obtained yet. Inspired by $|GHZ_n\rangle$  state contains information at $n$-party level and the properties of supports with its stabilizer formalism, we solve the problem of determination of stabilizer states. That is, in this paper, we proved that, among arbitrary states, the reduced density matrices for systems which
are the supports of $n$ independent generators of the corresponding stabilizer formalisms can uniquely determine the stabilizer states.

It is easy to see that the reduced density matrices for systems which are smaller than the supports of $n$ independent generators of the stabilizer can not determined the stabilizer states. We take the $4$-qubit graph state with the generators of its stabilizer are listed as Eq.(\ref{eq40}) for example. As we can see, the reduced density matrices set $\{{\rho ^{123}},{\rho ^{14}},{\rho ^{24}},{\rho ^{34}}\}$ can not determine this graph state, since the mixed state whose generators with the stabilizer  $\{ {\sigma _x} \otimes {\sigma _z} \otimes {\sigma _0} \otimes {\sigma _0},
 {\sigma _z} \otimes {\sigma _x} \otimes {\sigma _z} \otimes {\sigma _0},
 {\sigma _0} \otimes {\sigma _0} \otimes {\sigma _z} \otimes {\sigma _x}\}$ shares the same reduced density matrices with the graph state.
  We hope our paper can provide some insight into the characterization of multiparty entanglement with the stabilizer states.

\begin{acknowledgments}
This work is supported by NSFC (Grant Nos. 61272057, 61170270), Beijing Higher Education Young Elite Teacher Project (Grant Nos. YETP0475, YETP0477).
\end{acknowledgments}

\appendix
\section{}

We can get from the Eq.(\ref{eq29}) that
\begin{equation}
\label{a1}
\sum\limits_{{i_{{{\bar \omega }_s}}} \in \mathbb{F}_2^{|{{\bar \omega }_s}|}} {\left| {{b_{({i_{{\omega _s}}}{i_{{{\bar \omega }_s}}})({i_{{\omega _s}}}{i_{{{\bar \omega }_s}}} + {r_s})}}} \right|}  \ge \frac{1}{{{2^{|{\omega _s}|}}}}. 
\end{equation}
Since, for any $i,j$ it follows $\sqrt {{b_{ii}}{b_{jj}}}  \ge \left| {{b_{ij}}} \right|$,  we can get,
\begin{equation}
\label{a2}
\begin{split}
& \sum {\left| {{b_{({i_{{\omega _s}}}{i_{{{\bar \omega }_s}}})({i_{{\omega _s}}}{i_{{{\bar \omega }_s}}} + {r_s})}}} \right|}
  \\
  & \le \sum {\sqrt {{b_{({i_{{\omega _s}}}{i_{{{\bar \omega }_s}}})({i_{{\omega _s}}}{i_{{{\bar \omega }_s}}})}}{b_{({i_{{\omega _s}}}{i_{{{\bar \omega }_s}}} + {r_s})({i_{{\omega _s}}}{i_{{{\bar \omega }_s}}} + {r_s})}}} } \\
  & \le \sqrt {\left( {\sum {{b_{({i_{{\omega _s}}}{i_{{{\bar \omega }_s}}})({i_{{\omega _s}}}{i_{{{\bar \omega }_s}}})}}} } \right)\left( {\sum {{b_{({i_{{\omega _s}}}{i_{{{\bar \omega }_s}}} + {r_s})({i_{{\omega _s}}}{i_{{{\bar \omega }_s}}} + {r_s})}}} } \right)} \\
  &=\frac{1}{{{2^{{\omega ^s}}}}}. \\
\end{split}
\end{equation}
It follows from formulas (\ref{a1}) and (\ref{a2}) that all inequalities in these formulas should be equalities. Then for arbitrary  ${i_{{\omega _s}}} \in \mathbb{F}_2^{|{\omega _s}|}$ we have
\begin{equation}
\begin{split}
& \sum\limits_{{i_{{{\bar \omega }_s}}} \in \mathbb{F}_2^{|{{\bar \omega }_s}|}} {\left| {{b_{({i_{{\omega _s}}}{i_{{{\bar \omega }_s}}})({i_{{\omega _s}}}{i_{{{\bar \omega }_s}}} + {r_s})}}} \right|} \\
 & = \sum\limits_{{i_{{{\bar \omega }_s}}} \in \mathbb{F}_2^{|{{\bar \omega }_s}|}} {\sqrt {{b_{({i_{{\omega _s}}}{i_{{{\bar \omega }_s}}})({i_{{\omega _s}}}{i_{{{\bar \omega }_s}}})}}{b_{({i_{{\omega _s}}}{i_{{{\bar \omega }_s}}} + {r_s})({i_{{\omega _s}}}{i_{{{\bar \omega }_s}}} + {r_s})}}} }.\\ \nonumber
 \end{split}
\end{equation}
Therefore, for any vector $r \in \mathbb{F}_2^n$, it follows that
\begin{equation}
\left| {{b_{(r)(r + {r_s})}}} \right| = \sqrt {{b_{(r)(r)}}{b_{(r + {r_s})(r + {r_s})}}}. \nonumber
\end{equation}

From the Eq.(\ref{eq29}) and the following equation
\begin{equation}
\sum\limits_{{i_{{{\bar \omega }_s}}} \in F_2^{|{{\bar \omega }_s}|}} {\sqrt {{b_{({i_{{\omega _s}}}{i_{{{\bar \omega }_s}}})({i_{{\omega _s}}}{i_{{{\bar \omega }_s}}})}}{b_{({i_{{\omega _s}}}{i_{{{\bar \omega }_s}}} + {r_s})({i_{{\omega _s}}}{i_{{{\bar \omega }_s}}} + {r_s})}}} }  = \frac{1}{{{2^{{\omega _s}}}}}, \nonumber
\end{equation}
by employing the Schwartz inequality,  we can get
\begin{equation}
{b_{(r)(r)}} = {b_{(r + {r_s})(r + {r_s})}}. \nonumber
\end{equation}


\end{document}